\documentclass{article}

\usepackage{arxiv}
\usepackage[english]{babel}
\usepackage[utf8]{inputenc} % allow utf-8 input
\usepackage[T1]{fontenc}    % use 8-bit T1 fonts
\usepackage{hyperref}       % hyperlinks
\usepackage{url}            % simple URL typesetting
\usepackage{booktabs}       % professional-quality tables
\usepackage{amsfonts}       % blackboard math symbols
\usepackage{nicefrac}       % compact symbols for 1/2, etc.
\usepackage{microtype}      % microtypography
\usepackage{lipsum}
\usepackage{ dsfont }
\usepackage{chemformula}
\usepackage{siunitx}
\usepackage{authblk}
\usepackage{subfig}
\usepackage{array}
\usepackage{tabularx}
\usepackage{csquotes}

\title{Nanophotonic approaches for integrated quantum photonics}
\newcommand{\nanofiber}{NF}

\author[1,2]{Stefano Pierini}
\author[1]{Mackrine Nahra}
\author[2]{Maxime Joos}
\author[1,3]{Muhammad H. Muhammad}
\author[4]{Viatcheslav Agafonov}
\author[5]{Emmanuel Lhuillier}
\author[3]{Fabien Geoffray}
\author[6]{Valery Davydov}
\author[2]{Quentin Glorieux}
\author[2]{Elisabeth Giacobino}
\author[1]{Sylvain Blaize}
\author[2]{Alberto Bramati}
\author[1]{Christophe Couteau}

\affil[1]{University of Technology of Troyes and CNRS, Light, nanotechnologies \& nanomaterials-L2n, 12 rue Marie Curie, 10300 Troyes, France}
\affil[2]{Laboratoire Kastler Brossel, Sorbonne Université, Ecole Normale Supérieure and CNRS, 4 place Jussieu, 75252 Paris Cedex 05, France}
\affil[3]{TeemPhotonics, 61 chemin du Vieux Chêne
F-38240 Meylan, France}
\affil[4]{University of Tours, Greman, UMR CNRS CEA 6157, F. Rabelais University, 37200 Tours, France}
\affil[5]{Sorbonne Université, CNRS, Institut des NanoSciences de Paris, INSP, F-75005 Paris, France}
\affil[6]{L. F. Vereshchagin Institute for High Pressure Physics, Russian Academy of Sciences, Troitsk, Moscow 108840, Russia}
\begin{document}
\maketitle
\bibliographystyle{bibst/QUTE_bibstyle} 

\begin{abstract}
Photons for quantum technologies have been identified early on as a very good candidate for carrying quantum information encoded onto them, either by polarization encoding, time encoding or spatial encoding. Quantum cryptography, quantum communications, quantum networks in general and quantum computing~\cite{acin2018quantum} are some of the applications targeted by what is now called quantum photonics. Nevertheless, it was pretty clear at an early stage that bulk optics for handling quantum states of light with photons would not be able to deliver what is needed for these technologies~\cite{vamivakas2017special}. More recently, single photons, entangled photons and quantum optics in general have been coupled to more integrated approaches coming from classical optics in order to meet the requirements of scalability, reliablility and efficiency for quantum technologies. In this article, we develop our recent advances in two different nanophotonic platforms for quantum photonics using elongated optical fibers and integrated glass waveguides made by the so-called ion-exchange technique. We also present our latest results on quantum nanoemitters that we plan to couple and incorporate with our photonics platforms. These nanoemitters are of two kinds: nanocrystals made of perovskites as well as silicon-vacancy defect centres in nanodiamonds. Some of their properties are developed in this work. We will then give the general steps necessary in order to couple these nanoemitters efficiently with our platforms in the near future.
\end{abstract}

% keywords can be removed
\keywords{quantum photonics \and \ch{SiV} center \and perovskite nanocrystals}

\twocolumn

\section{Introduction}
For future quantum technologies and in particular for quantum architecture systems, one needs to consider scalability and reproducibility in order to be able to handle and add easily as many quantum bits/qubits of information as possible on a given platform. Quantum coherence and quantum information are fragile and thus require the need for many qubits to be created, to interact and quantum information to be preserved~\cite{divincenzo2000physical}. Currently there are either very good qubits in 
terms of fidelity but poor scalability or very good qubits in terms of scalability but poor fidelity thus requiring more qubits to compensate. Most of the current platforms are  in solid-state physics such as superconductors~\cite{krantz2019quantum}, dopants in silicon~\cite{kane1998silicon}, quantum dots in III-V semiconductors~\cite{loss1998quantum} or defects in diamond~\cite{childress2013diamond}. The main interest for condensed matter systems is the fact that it is potentially scalable as integration is possible in the future, as already demonstrated in the ‘classical’ semiconductor industry. 
Nevertheless, there are some true challenges in controlling and understanding the mechanisms of decoherence, losses and unwanted effects in these systems. As quantum technologies require ultimate control of quantum effects such as maintaining quantum coherence, quantum superposition and entanglement, it pushes towards deeper knowledge of underlying material and condensed matter physics. Photons are of particular interest as they are good carriers of quantum information and one aim is to explore a fully integrated photonic quantum circuit. This circuit would be a hybrid system made of stationary solid-state qubits (quantum emitters) coupled together via single photons travelling within a common optical bus.
This optical bus should be photonics-ready i.e. compatible with optical fibers for quantum communications within a network of quantum nodes. 
In this article, we will present our latest developments and results towards the integration of quantum emitters with photonic structures with nanosize to microsize features. In the first part, two platforms are described. 
The first one is based on the technique of elongated nanofiber~\cite{black1988tapered}
where the idea is to have an optical mode of the fiber mostly outside the fiber and as such will enhance the interaction with an outside emitter more efficiently. The second platform is made of glass and thus directly compatible with optical fibers. 
It is based on the technique of exchanged ions within the glass in order to create locally a particular confinement of the light \cite{tervonen_ion-exchanged_2011}. The next part will concern our latest results on two different novel quantum emitters. The first one is based on perovskite nanocrystals that can be synthetised chemically and giving rise to quantum optics-type of emitters. The second one is based on the so-called silicon-vacancy defect centre \ch{SiV} in nanodiamonds. 
These emitters are produced by the high-pressure/high-temperature method and have led to promising results for quantum technologies so far \cite{sipahigil2016integrated}. We will show that material science on the growth and post-processing is the key to understand and control the properties of these quantum emitters. Finally, the last part will describe our approaches in order to couple efficiently these platforms with these emitters towards integrated quantum photonics.

%\clearpage{}

\section{Photonic platforms under study}
\subsection{Nanofibers}
\label{sec:nanofibers}
%In the transition between two different linear materials, part of the inciding wave is reflected and part transmitted. The propagation radius of the transmitted part of the wave is deviated according to the Snell law, that can be written as $$ n_{1} \sin \theta_{1}=n_{2} \sin \theta_{2},$$ where the subscript $1$ and $2$ designs the two materials, $\theta$ is the angle formed with the normal to incidence plane and $n$ is the refraction index of the medium that describes the optical behaviour of the medium itself. In particular, moving from a more to a less refracting material (i.e. $n_1>n_2$) there are some values of $\theta_1$ for which this equation has no solution in $\mathds{R}$. In this case, the entire wave is reflected. 
%This proprierties is used to guide 

A nanofiber (\nanofiber) is a cylindrical glass optical waveguide with a diameter smaller than the wavelength of the guided light. It is created by heating and pulling a standard single mode fiber in order to reduce its diameter up to the desired size. At the end of the process, in the NF part, the core and the cladding of the fiber are merged together and the light is coupled at the interface between the \nanofiber{} and the air surrounding it. An adiabatic transition between the unpulled fiber and the \nanofiber{} and vice-versa allows a good light transmission of the whole system and an easy way to inject the light to and collect the light from the \nanofiber{}.
The choice of the nanofiber profile is crucial to obtain a good coupling and a good transmission and the techniques has been largely studied \cite{stiebeiner_design_2010,nagai_ultra-low-loss_2014,birks_shape_1992}.

In a \nanofiber{}, the fiber modes are outside the fiber and are very intense near the surface \cite{kien_field_2004}: this property allows the light to interact with atoms or nanoparticles placed in the proximity of the surface. At the same time, it allows the light emitted by the nano-object to be coupled to the fiber~\cite{nayak2008single}, with a precise relation between the polarization of the emitted light and the polarization coupled inside the fiber~\cite{joos2018polarization}. 

For the \nanofiber{} to be single mode, the following relation has to be valid: 
\begin{equation}
\label{eq:monomode}
V \equiv k a \sqrt{n_{1}^{2}-n_{2}^{2}}<V_{c} \cong 2.405
\end{equation}
where $a$ is the radius of the fiber, $n_1$ and $n_2$ are the two refractive index of the core and the cladding and $k$ is the wavenumber of the coupled light. In our case $n_2=1$, as in a \nanofiber{} the role of the cladding is played by the air surrounding it. 
In addition, in order to couple a nanoobject, we want the light intensity $\left|E\right|^2$ to be the strongest in the vicinity of the surface. This is obtained when the diameter of the \nanofiber{} is smaller then the light wavelength, typically by a factor 2.

\begin{figure}
    \centering
    \includegraphics[width=\linewidth]{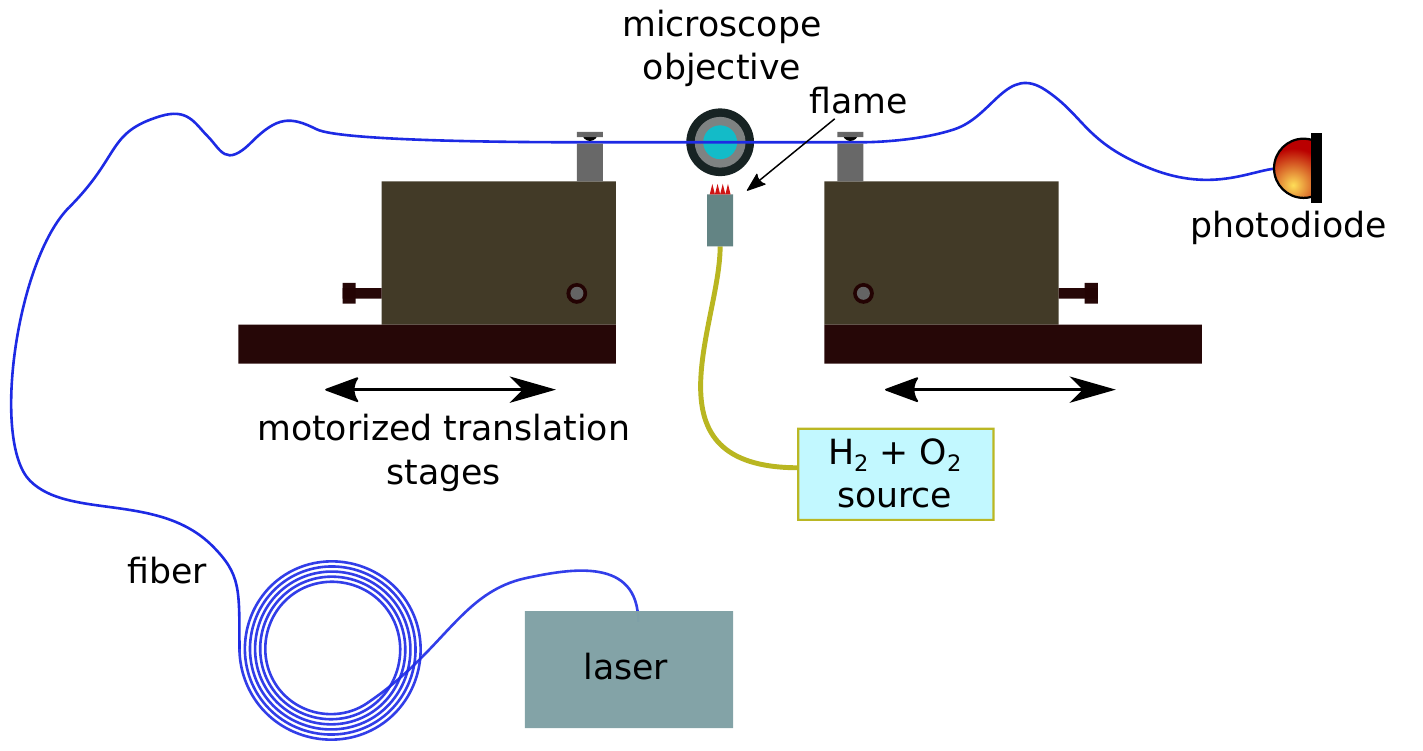}
    \caption{Scheme of the pulling setup. First the fiber is cleaned and clamped over the flame as shown, then the flame moves up and the two motorized platforms pull and move the fiber over the flame, while the procedure is imaged using a microscope objective on top. The transmission is monitored over the whole procedure by a photodiode at the output of the fiber. All the protocol is controlled by a computer.} 
    \label{fig:pulling_setup}
\end{figure}

The first step before to pull the fiber is to thoroughly clean it and remove the plastic jacket that covers the fiber. The fiber is then installed in the fabrication platform represented in figure~\ref{fig:pulling_setup}.
The fiber is suspended between two motorized platforms that can move horizontally and are controlled by a computer. A Bunsen burner is carefully positioned under the fiber and another motor controls its vertical position, and provide a controlled flame of \ch{H2} and \ch{O2}. A photodiode, placed at the output of the fiber allows to monitor the transmission during the whole pulling process. In addition, the image of the fiber is collected by a microscope objective and recorded by a camera behind it: this is useful to correctly position the flame and to precisely adjust the vertical position of the flame during the pulling. 
When the pulling starts, the flame is placed at a given distance from the fiber and the motors start moving and pulling the fiber in order to reach the given shape. 
The movements needed to perform this operation depend not only on the chosen profile, but also on the physical characteristics of the flame, such as its size and temperature. The distance between the flame and the fiber, as well as the flux of hydrogen and oxygen need to be carefully chosen in order to obtain the same effective flame diameter used to calculate the profile itself, in our case \SI{0.5}{\milli\meter}. 
The effect of the flame size in the pulling procedure is described in~\cite{hoffman2014optical}.
With this system, we were able to reach fiber radius as small as \SI{150}{\nm} and to reach transmissions over $95\%$, but by accurately optimizing the shape of the tapered part even higher transmissions can be achieved~\cite{nagai_ultra-low-loss_2014}.

In order to deposit a nanoemitter over the fiber, we create a drop of solution of \SI{20}{\micro\liter} containing the nanoemitters at the end of a micro-pipette. The next step is then to touch the \nanofiber{} with the drop repetitively until a particle gets stuck to it. For more precision, the whole process is controlled with a microscope and the micro-pipette is moved via precise translation stages.

\paragraph{Experimental apparatus}

\begin{figure}
    \centering
    \includegraphics[width=\linewidth]{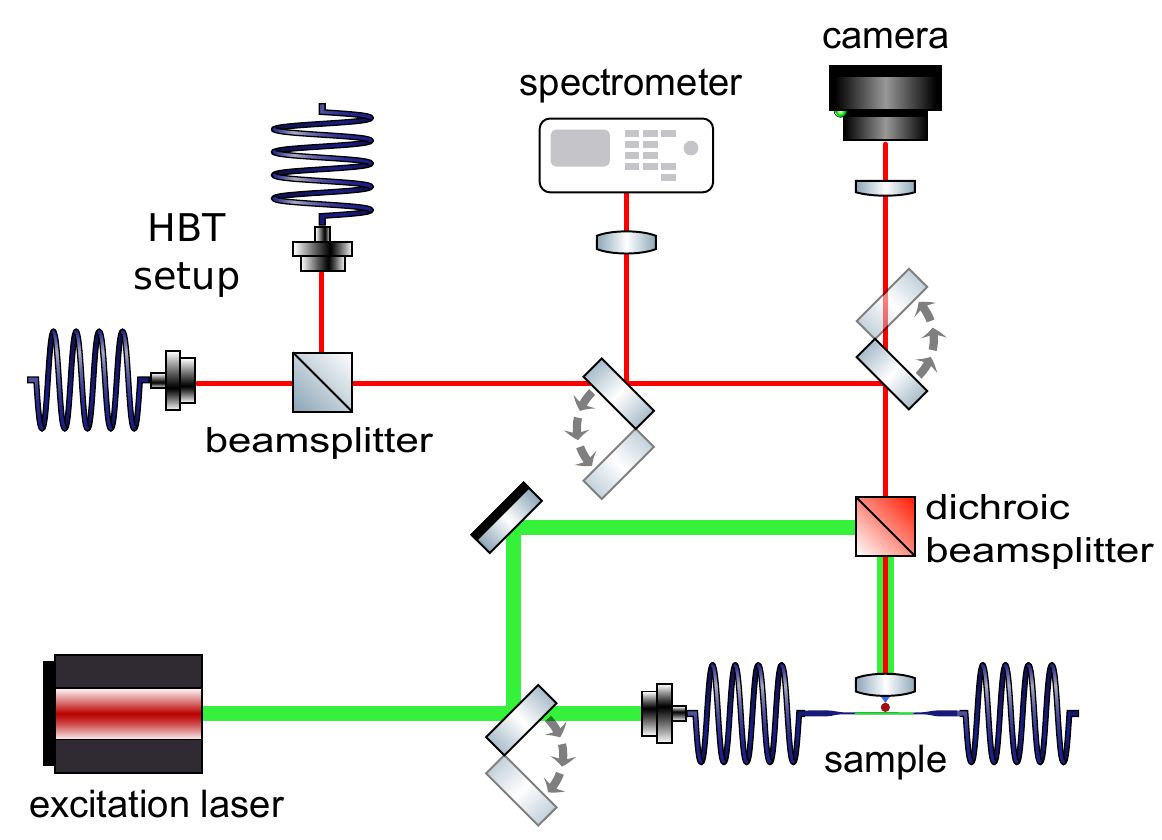}
    \caption{Scheme of the optical setup used to measure the light emitted by nanoemitters deposited on the \nanofiber{}. The green line show the path of the excitation laser (whom wavelength depends on the emitter itself). Changing the position of the beamsplitter we can send it through our confocal microscope or directly through the \nanofiber{}. The red line shows the detection path, that allows to measure alternatively the image of the emitter, its spectrum or the $g^2(\tau)$ function.}
    \label{fig:fiber_setup}
\end{figure}

Our setup is represented in figure~\ref{fig:fiber_setup}. By sending the light in the fiber we can excite the emitter and see the emission through the microscope: this is very convenient to detect the emitted light without having to care about the precise position of the emitter and to verify its single photon emission. If the coupling is good enough, we can also excite the nanoemitter through the microscope and collect the light from the fiber.

First results were obtained on our set-up with semiconductor nanocrystals that are well known to be good single photon emitters~\cite{michler2000quantum} and so well adapted for first tests. 
The emitters we used are dot-in-rod \ch{CdS/CdSe} nanocrystals: this kind of nanocrystals have a core/shell structure which ensures a reduced blinking of the emission;
moreover dots in rod have the specificity to have elongated shell that makes the emission to be polarized~\cite{pisanello2010room}.

\begin{figure}
    \centering
    \includegraphics[width=\linewidth]{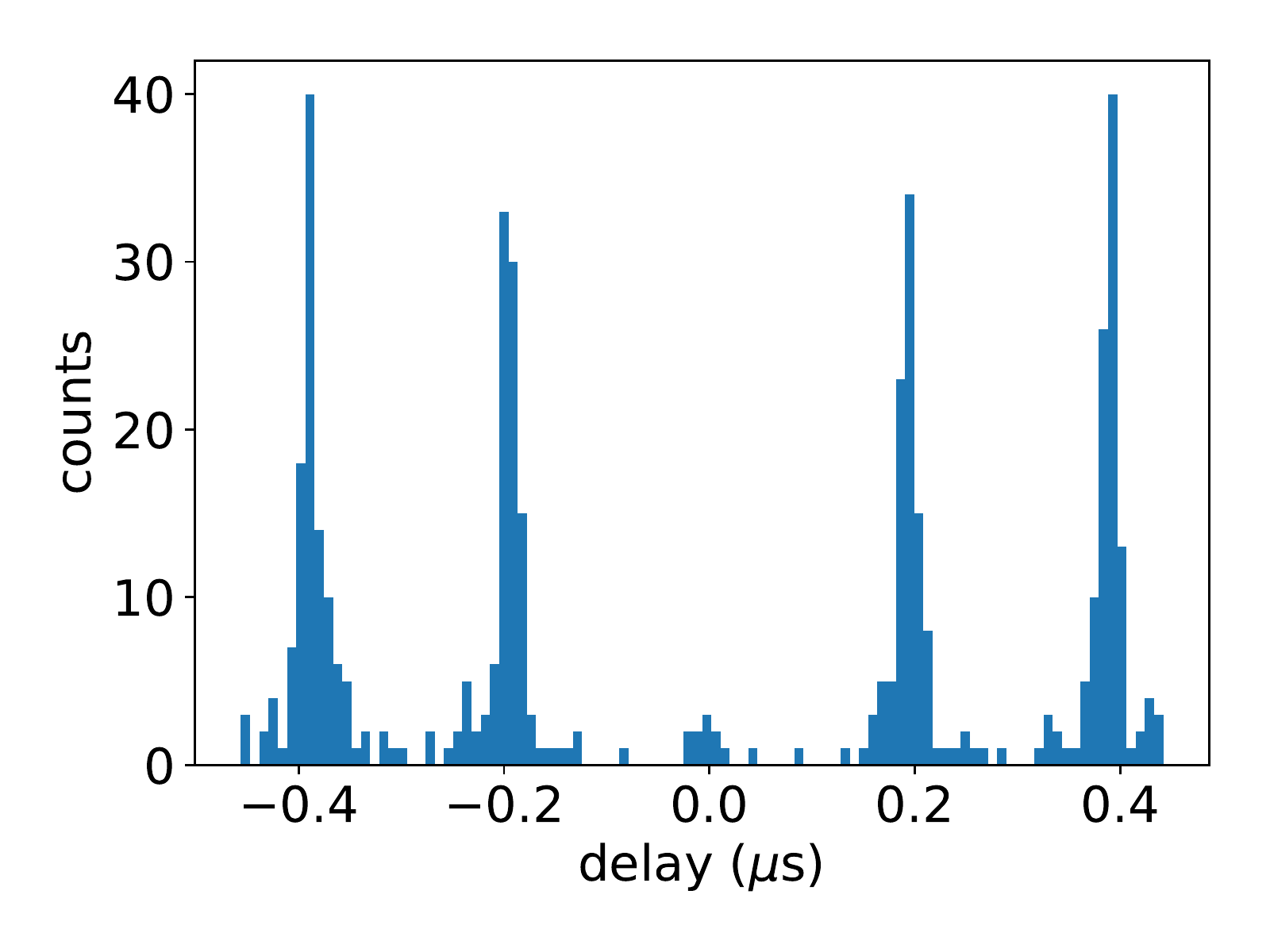}
    \caption{$g^{(2)}(\tau)$ function measured by exciting a dot-in-rod nanocrystal deposited on the \nanofiber{} with a \SI{405}{\nm} pulsed laser. The repetition rate was \SI{5}{\mega\hertz}. The photons were collected via the \nanofiber{}. }
    \label{fig:g2_from_fiber}
\end{figure}

For that, we pulled a fiber with a diameter of \SI{300}{\nm} in order to guide the \SI{600}{\nm} light emitted by the nanocrystals. As a result, we were able to excite a single nanocrystal from the free space. The emitted light collected by the \nanofiber{} was enough to measure the photon autocorrelation function ($g^{(2)}(\tau)$) using a Hanbury Brown and Twiss set-up (figure~\ref{fig:g2_from_fiber}). For this experiment a pulsed excitation laser with a wavelength of \SI{405} {\nm} was used. A clear signature of a single photon emitter was observed with a $g^{(2)}(\tau)$ smaller than $0.2$. This first result validates our experimental protocol for the use of other emitters and confirm that we can use different emitters in future experiments such as perovskite nanocrystals or silicon-vacancy centers in nanodiamonds as described further down.

\subsection{Optical glass waveguides}
\begin{figure}
    \centering
    \subfloat[][\label{fig:Wave_tph_a}]{
    \includegraphics[width=0.45\linewidth]{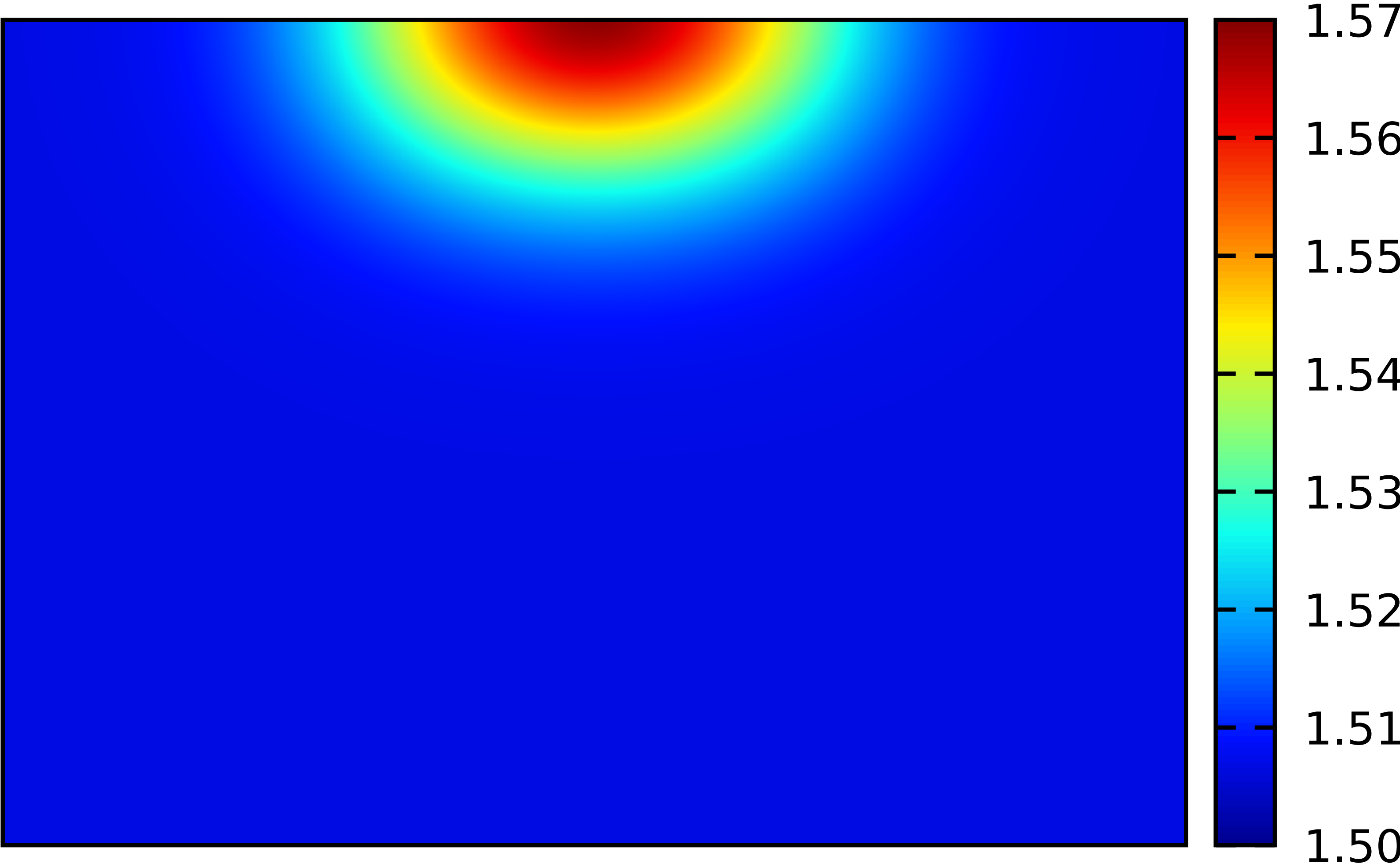}}
    \quad
    \subfloat[][\label{fig:Wave_tph_b}]{
    \includegraphics[width=0.45\linewidth]{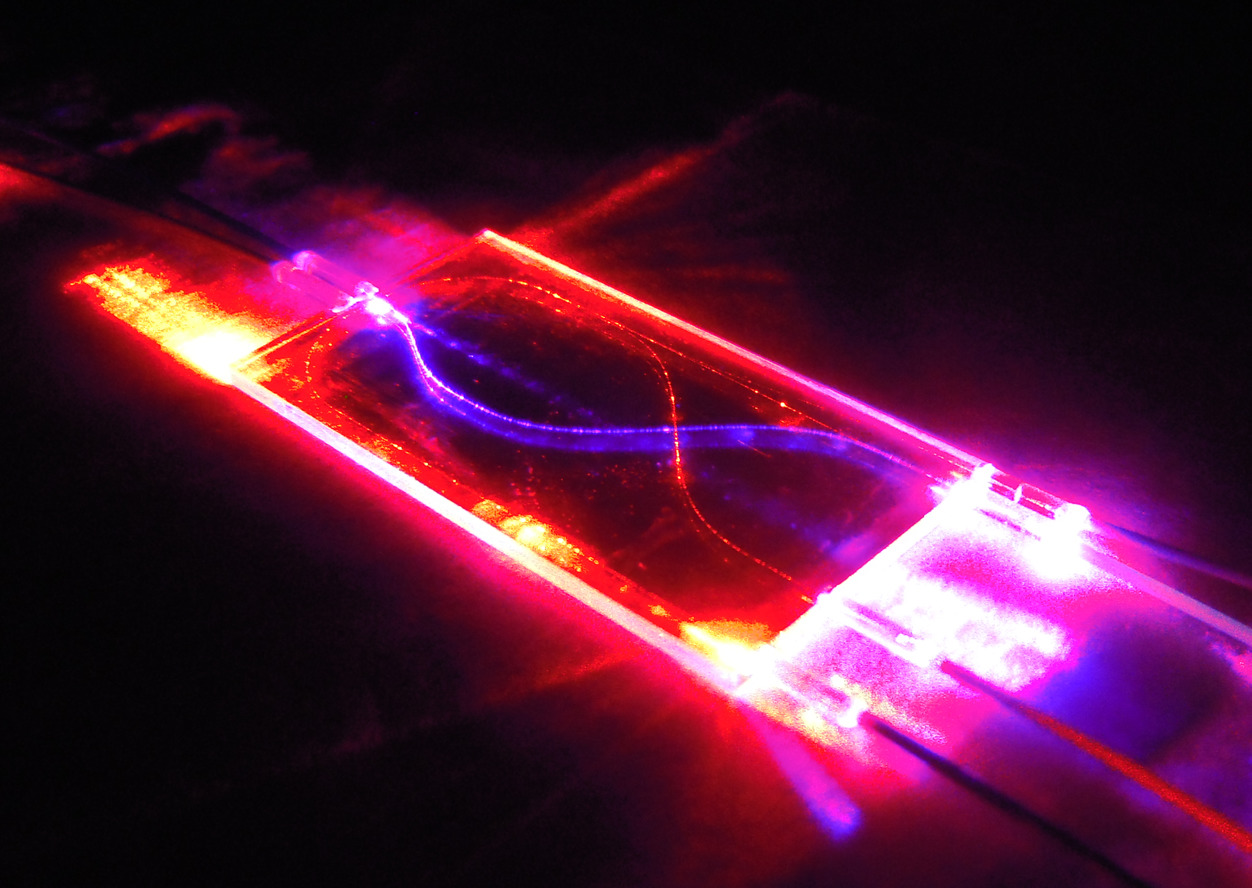}}
    \caption{\protect\subref{fig:Wave_tph_a} Typical gradient index profile of a surface waveguide made by the ion-exchange technology. \protect\subref{fig:Wave_tph_b} Example of a beam combiner glass chip working in the visible wavelength range.}
    \label{fig:Wave_tph}
\end{figure}

%(Muhammad + Fabien) 1 page & 1 to 2 figures
The notion of integrated optics was first introduced back in 1969 by S.~E.~Miller~\cite{miller1969integrated} with the revolutionary idea to reuse the planar technologies developed at the time for the microelectronics in order to fabricate optical circuits. 
Such devices opened up the way to compact, robust and self-aligned optical systems. Since the early years, the glass platform raised a great interest for many reasons.
Indeed, glass waveguides show very low propagation losses – on the order of 0.1dB/cm – on a large bandwidth from visible to near-infrared wavelength, while ensuring a high coupling efficiency with optical fibers. In our case, these waveguides are realized on glass thanks to the ion-exchange technology known for centuries (with stained glasses) to change the glass color properties~\cite{mazzoldi2013ion}. 
The waveguide core creation relies on the diffusion – either purely thermal or field assisted – of alkali ions into the glass substrate which induces in turn an increase of the refractive index. Different ionic species (\ch{Tl+}, \ch{K+}, \ch{Ag+}) can be exchanged with the one hosted in the glass matrix (\ch{Na+} being the most common one)~\cite{broquin2001ion} 
resulting in a gradient-like profile. In order to make strip waveguides, the ion-exchange process is limited to a specific area using a photo-lithography technique. An example of a simulated index profile obtained for a 2$\mu$m mask opening is depicted in figure~\ref{fig:Wave_tph_a}.
Such surface waveguide is designed to work in single mode operation at \SI{1550}{\nano\meter} telecommunication wavelength. 
Figure~\ref{fig:Wave_tph_a} shows the local increase of the index which is characterized by a low refractive index contrast, $\Delta \textrm{n} = 0.052$ here. The first demonstration of an ion-exchanged waveguide (IEW) was achieved in 1972~\cite{izawa1972optical} and the technology has been continuously investigated and developed since then~\cite{tervonen_ion-exchanged_2011}. 
As such, the glass optical platform is highly versatile and can offer both surface and buried waveguides designs as opposed to other techniques such as the laser-writing technique. For this technique, surface waveguiding is not possible and thus interaction with nanoemitters from the top is not possible. Borosilicate glasses are commonly used to implement passive features and to make complex circuitry. For instance a beam combiner implemented on a glass chip is illustrated in figure~\ref{fig:Wave_tph_b}. 
The IEW are connected to optical fibers thanks to glass ferules as visible on the bottom right of figure~\ref{fig:Wave_tph_b} ensuring a good coupling efficiency. Nowadays, optical glass waveguides are widely spread for many applications including telecommunication and
sensing~\cite{tervonen_ion-exchanged_2011}. 
For instance, chemical analysis takes benefit of the evanescent field interaction of a surface waveguide~\cite{allenet2019microsensing} to perform absorption spectroscopy in harsh environment. The latter phenomenon can also be used to couple the light from a nanoemitter~\cite{madrigal2016hybrid} for quantum applications. This is precisely the effect that we want to use for coupling nanoemitters (such as the ones described below) to these optical buses for photons.

\section{Quantum Emitters}
\subsection{Perovskites nanocrystals}
\begin{figure}
    \centering
    \includegraphics[width=\linewidth]{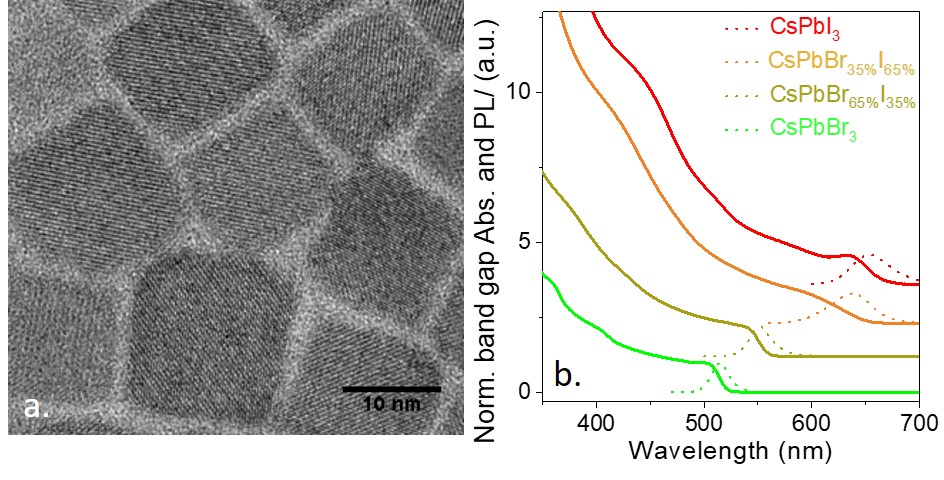}
    \caption{a) transmission electron microscopy image of \ch{CsPb(Br ;I)3} nanocrystal cubes. b) absorption and photoluminescence of \ch{CsPb(Br ;I)3} perovskites nanocrystal with various \ch{Br} content.}
    \label{fig:perov1}
\end{figure}

\paragraph{General description}
Nanocrystals are semiconductor nanoparticles with broadly tunable optical features from UV to THz \cite{goubet2018terahertz}. Thanks to quantum confinement, atomic-like spectra can be obtained from such objects \cite{murray1993synthesis}. They thus appear to be an interesting candidate for quantum optics as single photon emission has been reported from them in the past \cite{brokmann2004highly}. Nevertheless,  traditional II-VI semiconductor nanocrystals suffer from several issues. In particular, bright emission is only obtained through the growth of core shell heterostructures. Meanwhile, perovskite nanocrystals made of lead halide materials became very popular in the field of solar cells. Their defect-tolerant electronic structure leads to improved open-circuit voltage. It has been proposed to take advantage of this defect tolerance of lead halide perovskite to design bright core only nanocrystals \cite{Protesescu2015}. 
The basic concept of the synthesis relies on the reaction of lead halide, in presence of long chain ligands made of oleic acid and oleylamine in a non coordinating solvent. At relatively low temperature (180 °C), the injection of a cesium oleate precursor leads to an immediate formation of \ch{CsPbX3} nanocrystals, with \ch{X=Cl}, \ch{Br} or \ch{I}  with cubic shape: this is clearly visible in figure \ref{fig:perov1}a where a transmission electron microscopy image of \ch{CsPb(Br ;I)3} is shown.
%\textbf{(describe figure more)}
These particles have size around \SI{10}{nm} which is higher than the Bohr radius meaning their behaviour is very close to the bulk material. 
As for II-VI semiconductor nanocrystals, the band-edge energy can be tuned thanks to quantum confinement. Smaller nanocrystals can simply be obtained by reducing the growth temperature. However this is not the method the most commonly used. Conventional alloying is usually used. We can see in figure~\ref{fig:perov1}b
 the  absorption and photoluminescence spectra of different alloying perovskite-nanocrystals in solution leading to different colors.
%, see figure \ref{fig:perov1}b. 
The whole visible range can be spanned by tuning the X from \ch{Cl} to \ch{I}. The most striking property of these perovskite nanocrystals comes from their high photoluminescence efficiency which ranges from $50\%$ to $90\%$ depending on the chosen halide. The investigation of the single particle property of \ch{CsPbX3} nanocrystals can easily be obtained by diluting them into a polystyrene matrix~\cite{raino2019underestimated}.

\paragraph{Optical characterization} 
\begin{figure}
    \centering
    \includegraphics[width=\linewidth]{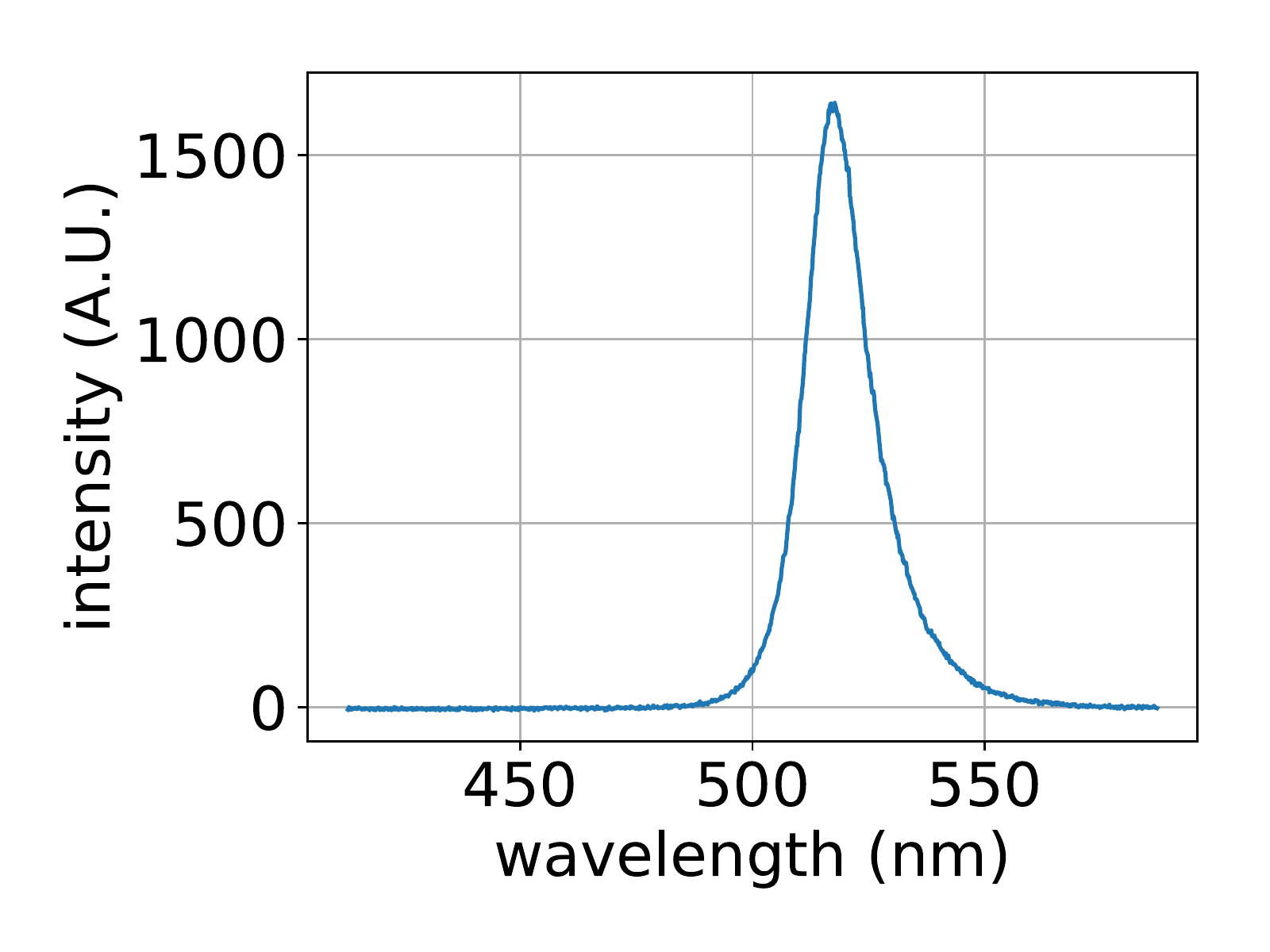}
    \caption{Spectrum of a perovskite nanocrystal with a center wavelength of \SI{518}{nm} and a full width at half maximum \SI{16}{nm}.}
    \label{fig:perov_spectra}
\end{figure}
To study the emission spectra of these nanocrystals, we deposite them on a glass plate and use a confocal microscope to excite a single emitter with a \SI{405}{nm} pulsed excitation laser. The center wavalenghts of the spectra spread from \SI{480}{nm} to \SI{520}{nm} with a peak with a full width at half maximum (FWHM) of about \SI{15}{nm}. A typical spectrum is shown in figure \ref{fig:perov_spectra}.
\begin{figure}
    \centering
    \includegraphics[width=\linewidth]{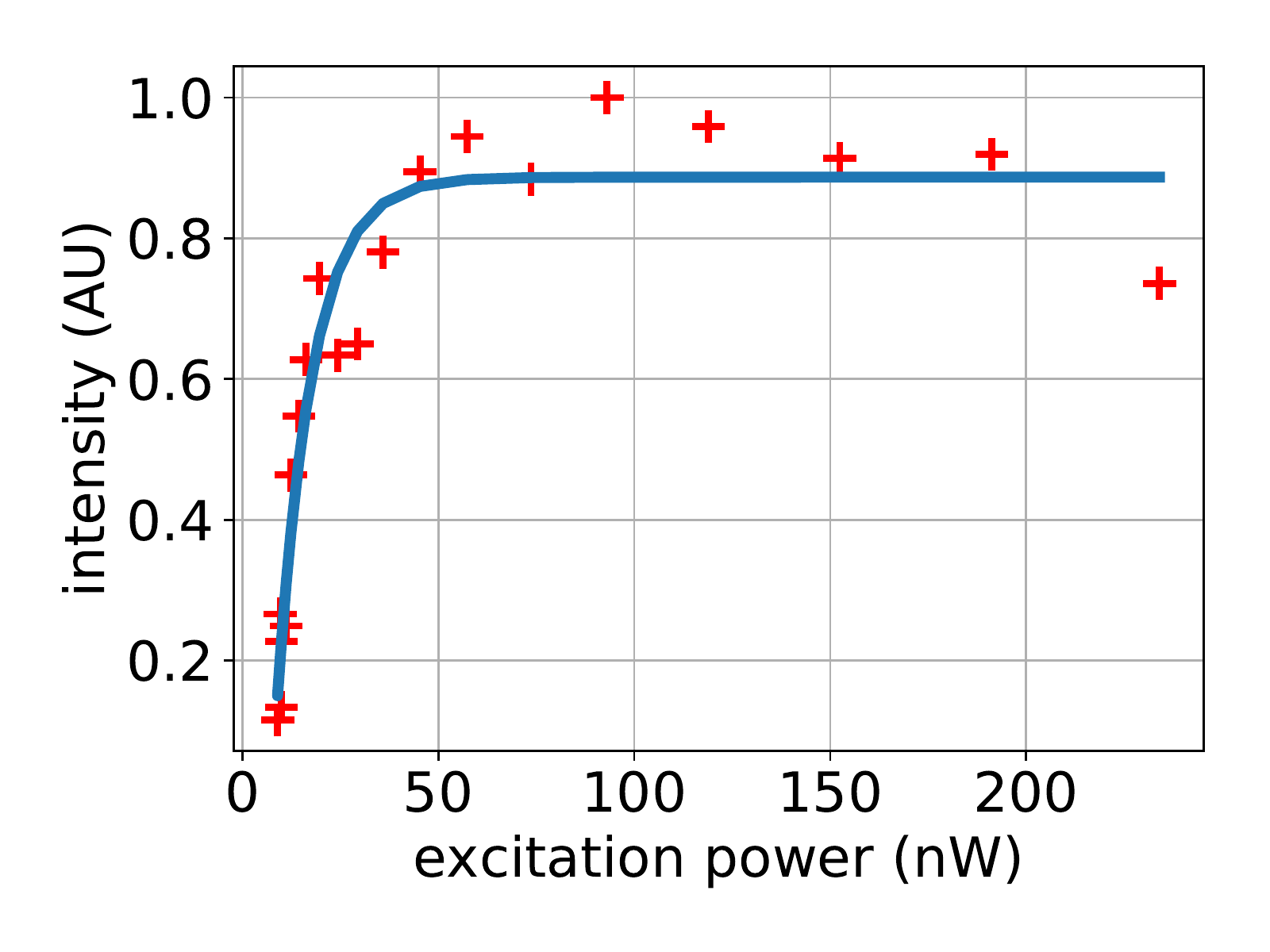}
    \caption{Saturation curve of a perovskite nanocrystal showing single photon emission.}
    \label{fig:perov_saturation}
\end{figure}
The emission of a single nanocrystal shows a clear saturation of the emitted light while increasing the excitation power, with a saturation power of \SI{80}{nW} (the curve is shown in Figure~\ref{fig:perov_saturation}). Each experimental point is obtained by taking ten measurements and keeping only the three brightest ones. This procedure reduce the effect of blinking providing a most reliable result.
The measurements of the Stokes parameters performed on our samples with the method illustrated in
\cite{berry1977measurement} have shown that the emission is not polarized.   

\paragraph{Quantum properties} Antibunching measurements were performed using a Hanbury Brown and Twiss setup showing that some of the emitters had a clear single photon emission, with a $g^{(2)}(0)<0.1$. 
With our setup we can measure the $g^{(2)}(\tau)$ 
for long values of $\tau$ (up to hundreds of \si{\micro\second}). This feature allows us to normalize the $g^{(2)}$ function at large delays, taking into account the bunching effect due to the blinking. This effect has to be considered in order to identify the real quality of the emitter and it is taken in consideration for the first time for perovskite nanocrystals.
\begin{figure}
    \centering
    \includegraphics[width=\linewidth]{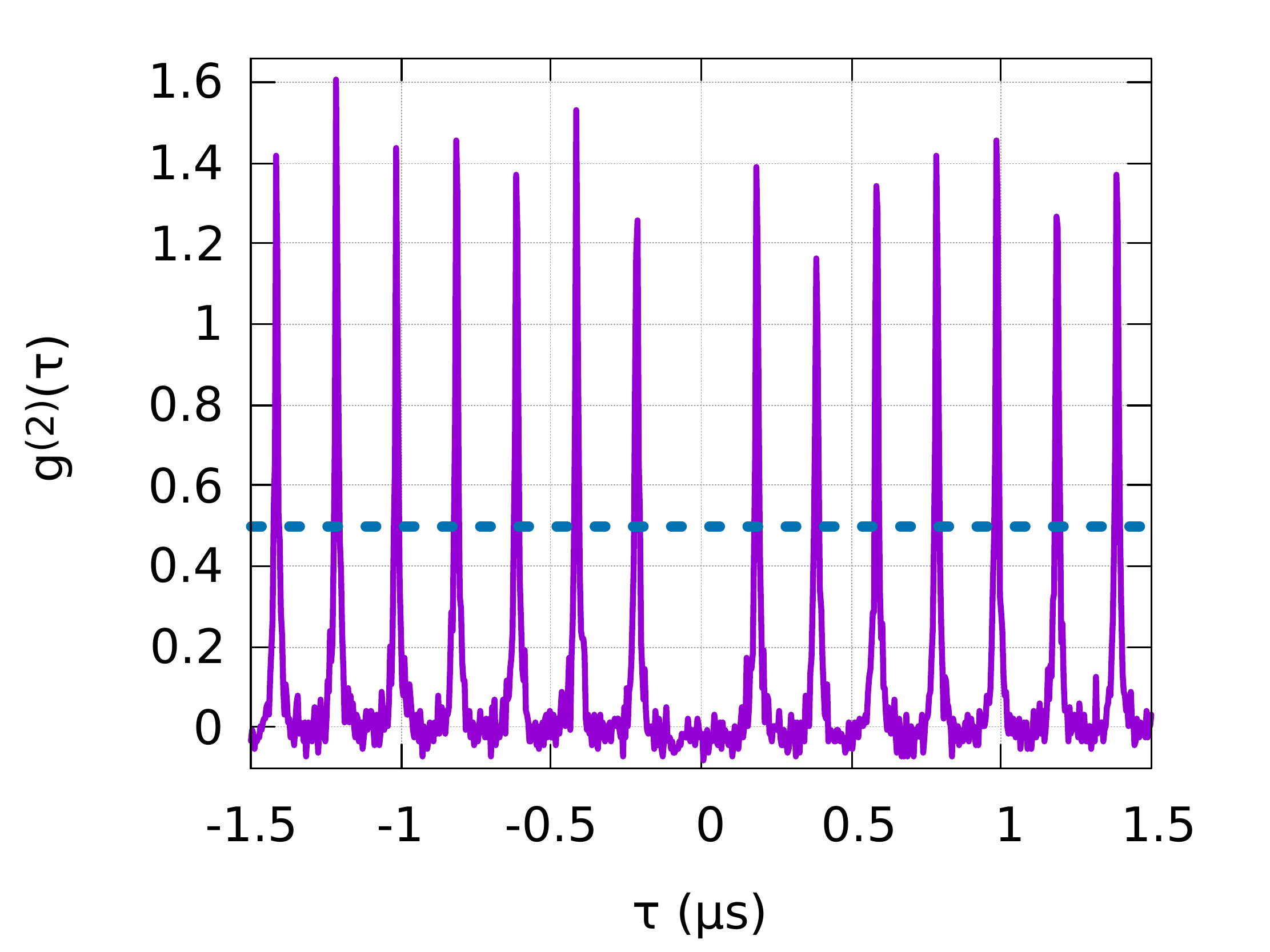}
    \caption{measurement of  the autocorrelation function $g^{(2)}(\tau)$ of a single perovskite nanocrystal, showing clear single photon emission. The data are corrected for the noise (background removed) and normalized at large delays ($\tau \approx \SI{100}{\micro\second}$).}
    \label{fig:g2_perov}
\end{figure}

The result of the photon antibunching is shown in figure~\ref{fig:g2_perov}. The experimental curves are normalized using the peaks at a long delay
$\tau\approx\SI{100}{\micro\second}$, 
after having subtracted the background noise, estimated using the values between two different peaks. To perform the measurement, we use a PicoHarp~800 with a router that registers the full history of the detected photons. There is a limitation as two events occurring with a delay smaller than \SI{100}{\nano \second} cannot be registered and thus creating an artificial flat region that has been removed from the signal for more clarity. A blue line indicates the $0.5$ threshold under which the emitter can be regarded as a single photon one. We can also note that the blinking of the emitter creates a bunching effect, that is clearly visible as the other peaks are higher than $1$: this kind of effect was previously observed also with other types of emitters~\cite{manceau2018cdse, ulrich2005correlated}.  

\subsection{\ch{SiV} defect centers in nanodiamonds}
\subsubsection{Silicon-vacancy defect centers}
Color center defects in diamond have become one of the leading research subjects in recent years due to their attractive optical properties. 
Among these, silicon vacancy~(\ch{SiV}) centers are very promising thanks to their attractive features, such as high brightness \cite{neu2011single}, narrow inhomogeneous distribution \cite{sternschulte19941}, stable single photon emission (even at room temperature) and minimal spectral diffusion \cite{rogers2014multiple}. 
These properties of \ch{SiV} are desirable for quantum photonics especially for quantum information \cite{prawer2014quantum}, quantum communication \cite{su2009high} and quantum networks \cite{sipahigil2016integrated}.
Bulk diamond induces a low extraction of single photons from \ch{SiV} centers due to its high refractive index ($n_d=2.41$). To overcome this problem, diamond nanostructures have been adopted. They can be spatially manipulated and positioned to enhance coupling to other nanophotonics structures or to fibers.  
Here, we explore the possibility of controlling the nanodiamonds size distribution and photoluminescence of \ch{SiV} centres by varying growth conditions. 
We perform room-temperature characterization of several ND samples grown using the high pressure/high temperature (HPHT) method \cite{davydov2014production}. 
Previously, it has been found that nanodiamonds grown with this method host \ch{SiV} centers with excellent optical properties \cite{rogers2019single}. 
They are, however, limited by effects due to the surface of the nanodiamonds. 
In this article, we explore this issue by studying the influence of the growth temperature and the treatment of the surface with mixture of acids on the physical and optical properties of our nanodiamonds.
Table 1 shows the growth conditions for different nanodiamonds samples A, B and C.

\begin{figure*}
    \centering
    \subfloat[][Sample A\label{fig:sampleA}]{
        \includegraphics[width=0.32\linewidth]{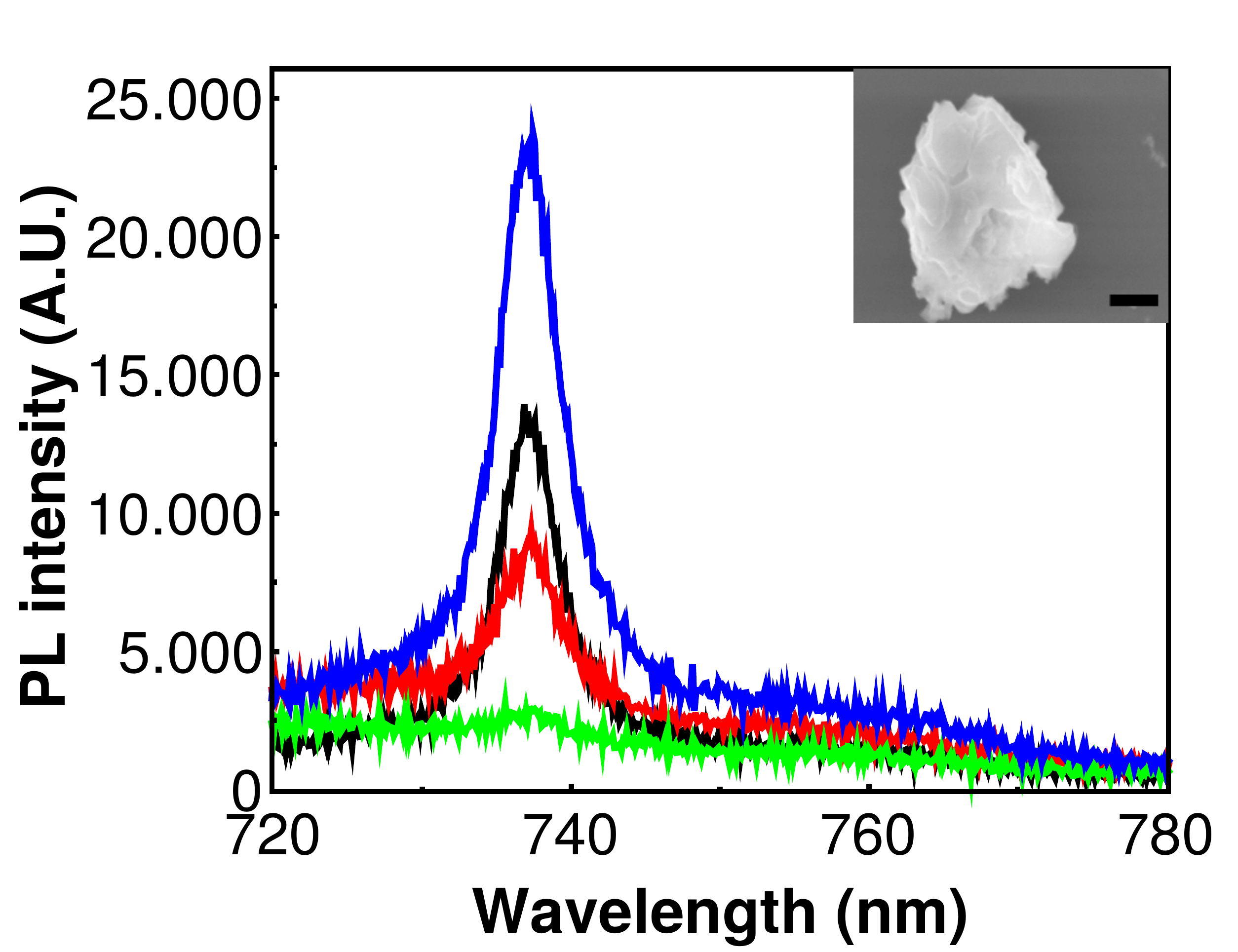}}
    \subfloat[][Sample B\label{fig:sampleB}]{
        \includegraphics[width=0.32\linewidth]{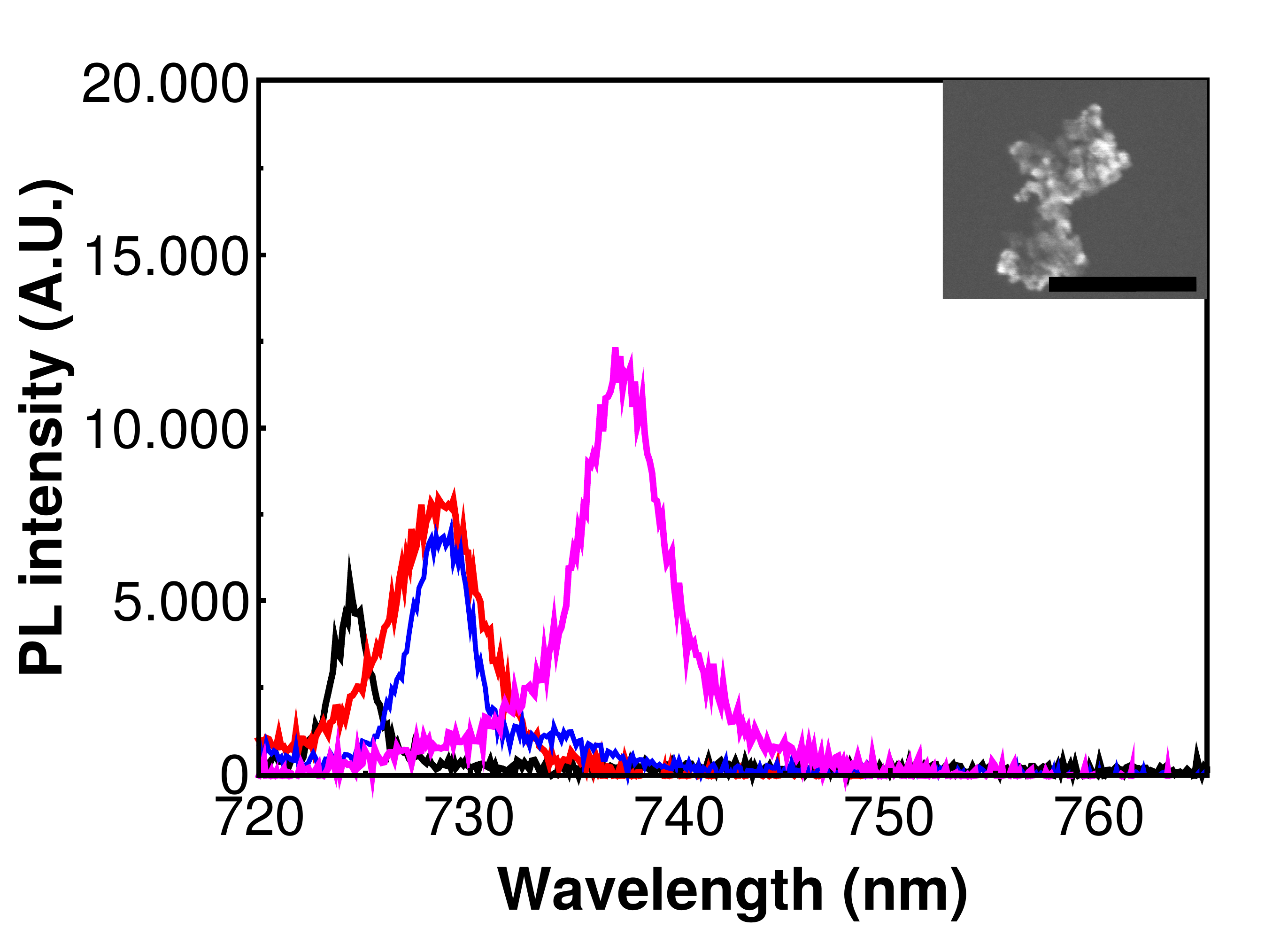}}
    \subfloat[][Sample C\label{fig:sampleC}]{
         \includegraphics[width=0.32\linewidth]{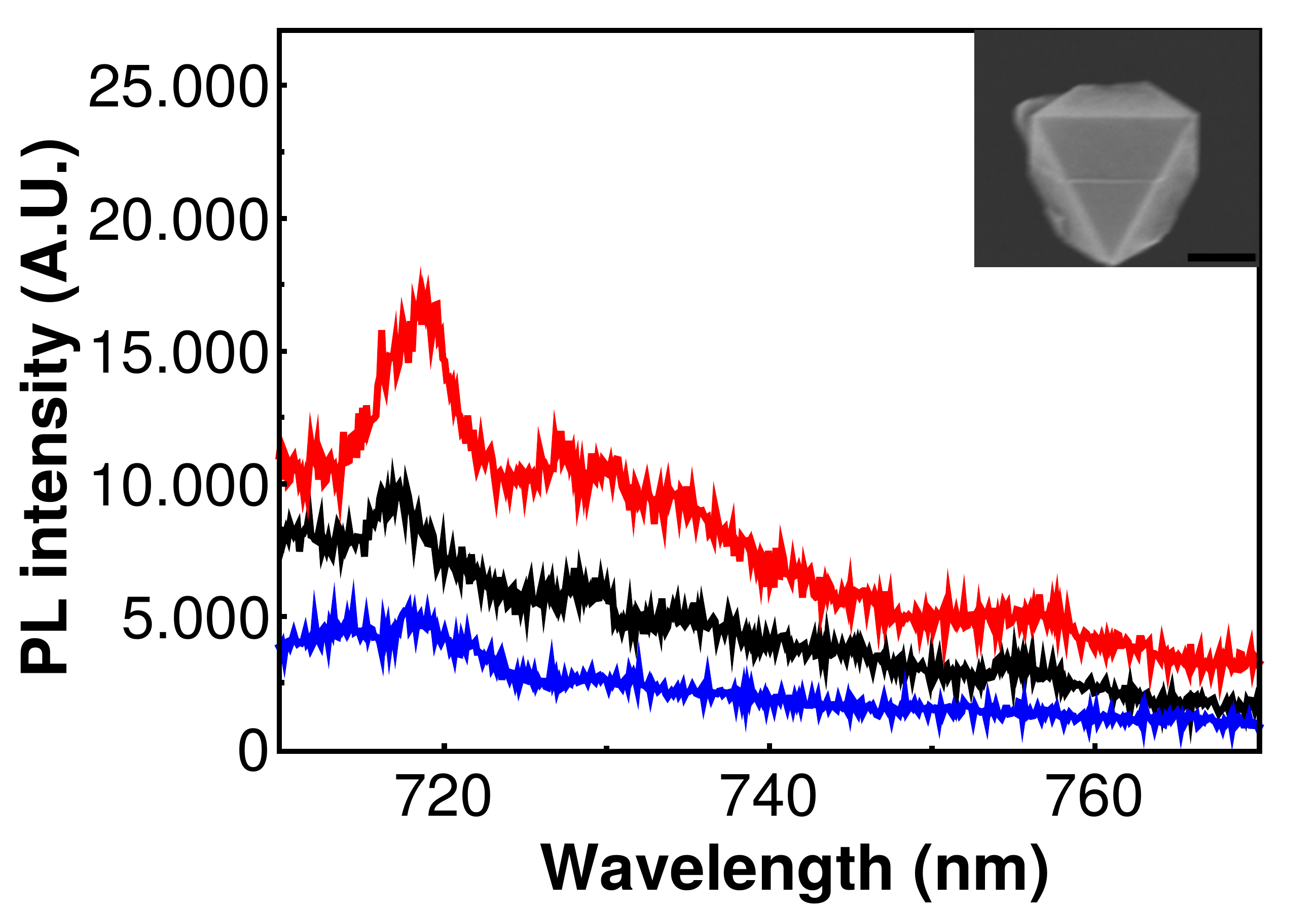}}
    
    \caption{Photoluminescence spectra of three differently grown \ch(SiV) nanodiamonds with insets showing their respective SEM pictures.  Samples A, B, and C refers to the growth conditions reported in Table~\ref{tab:table}. A clear difference of their emission and morphology is shown. Black lines in these insets represent a scale of 200 nm.}
    %\caption{Photoluminescence spectra of samples A, B, and C with insets showing their respective SEM pictures. A clear difference of their emission and morphology is shown. Black lines in these insets represent a scale of 200 nm.}
    \label{fig:PL_SiV}
\end{figure*}

\subsubsection{Growth of nanodiamonds containing SiV centers}
The NDs with \ch{SiV} centers used in this work were synthesized using the same high-pressure high-temperature (\ch{8}{GPa}) method described in ref~ \cite{davydov2014production}, based on mixtures of naphthalene (\ch{C10H8}), \ch{CF1,1} (fluorographite) and tetrakis(trimethylsilyl) silane (\ch{C12H36Si5}, Strem Chemicals Co.) was used as the main silicon doping addition. 
The samples studied here are different in terms of growth temperature and surface treatment.  
Two temperatures were studied here, \SI{1350}{\celsius} (B) and \SI{1500}{\celsius} (A and C), to explore the effect on the size of the NDs. 
In addition, acid treatment is studied here to explore its effect on the shape and the crystallinity of the nanodiamonds (samples A and C).
\begin{table}[h]
\begin{tabular}{cccc} \toprule
Sample & Si/C ratio &Temperature ($^{\circ}$C) & Treatment \\ \midrule
 A & $0.079$& $1500$ & Acid \\
 B & $0.01$ & $1350$ & Acid \\
 C & $0.007$& $1500$ & -----\\ \bottomrule
\end{tabular}
\caption{Growth parameters of nanodiamonds samples}
\label{tab:table}
\end{table}

\subsubsection{Experimental setup}
To analyze the optical properties of the SiV centers in nanodiamonds, we use a home-built microphotoluminescence set-up at room temperature, as described in the following.A He-Ne laser with a wavelength of \SI{632.8}{nm} excites the sample through an air microscope objective ($\textrm{NA}=0.95$, $\times100$). The emission passes through a high pass filter (above 700 nm) so that the excitation light is rejected. It is then directed to either a grating spectrometer (\SI{0.07}{nm} resolution) for photoluminescence measurements or to a multimode fiber (\SI{65}{\micro \meter} core) that acts as a pinhole. In the second case, the emission is sent afterwards to a Hanbury Brown and Twiss setup. The latter consists of a 50/50 beam-splitter and two avalanche photodiodes (APD) that are used for single photon counting to measure the $g^{(2)}(\tau)$ second-order correlation function. Finally, a narrow-band filter (13 nm bandwidth around 740 nm) is placed in front of each APD. The purpose of these filters is double: i- they remove all background emission and ii- they eliminate the cross talk between the APDs.

\subsubsection{Emission properties of the SiV centers}
\paragraph{Optical properties: photoluminescence}
Photoluminescence measurements of stable \ch{SiV} centers in sample A (\SI{1500}{\celsius}) revealed fluorescence peaks at the typical wavelength with a zero phonon line (ZPL) at \SI{737}{nm} and with a visible phonon sideband (see Figure~\ref{fig:sampleA}).
As for SiV defects in sample B ( \SI{1350}{\celsius}), they suffer from surface effect from a highly strained nanodiamonds, revealed in an inhomogeneous distribution of the ZPL varying between \SI{727}{nm} and \SI{740}{nm}.(see Figure~\ref{fig:sampleB}).
Similar observation of local strain in smaller  nanodiamonds were reported elsewhere \cite{neu2013low} \cite{neu2011single}. 
The tiny size of nanodiamonds is coming from the fact of lower growth temperature.

\paragraph{Physical properties: morphology}
Figure~\ref{fig:sampleC} presents a highly crystalline nanodiamond under our electron microscope and yet, despite the crystalline quality of the nanodiamonds in sample~C , no emission of \ch{SiV} centers was found. The PL spectra show a clear signature of graphite though (peak at \SI{720}{nm}). We deduce that the non-treatment of sample C with acid favours other defects and impurities at the surface, such as surface contamination and graphitization which induces the quenching of luminescence. 
Sample A was grown at the same temperature but was subjected to acid treatment and clearly presents nice photoluminescence at \SI{737}{\nano\meter} or so Figure~\ref{fig:sampleA}. 
Such acid treatment is thus needed to prevent the contamination and graphitization on the surface of nanodiamonds.

\section{Towards coupling of nanoemitters using nanophotonics}

\begin{figure}[b]
    \centering
    \includegraphics[width=\linewidth]{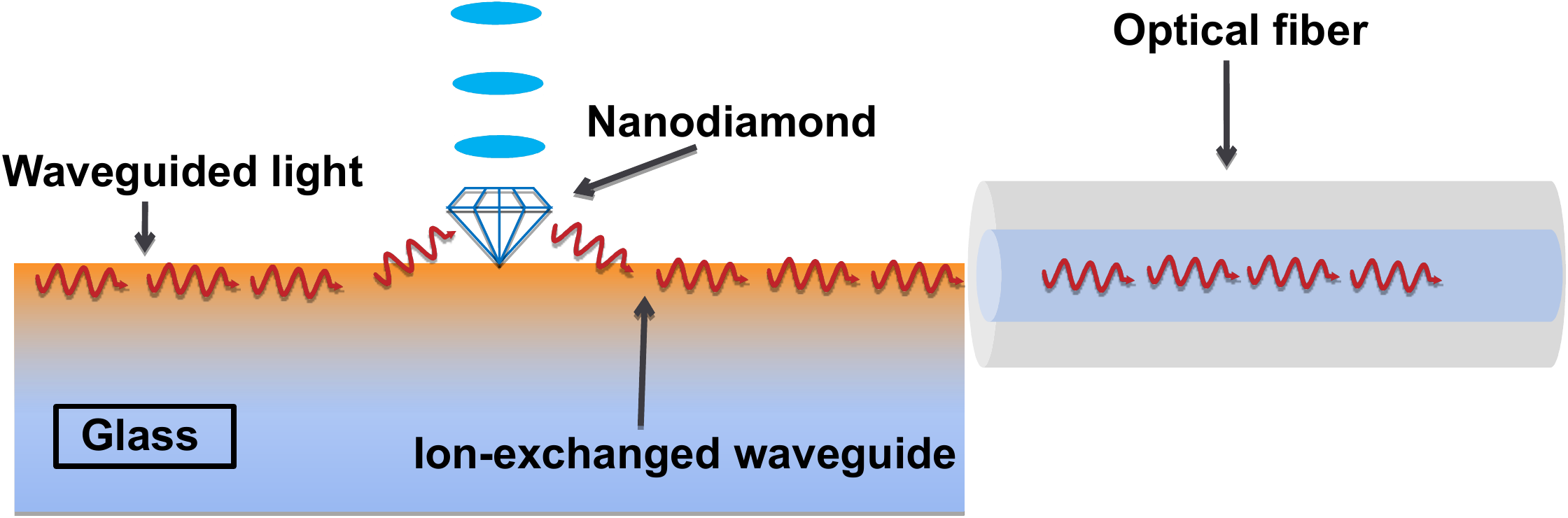}
    \caption{Schematic of a single nanodiamond coupled to an ion-exchange waveguide, itself coupled to an external optical fiber. Excitation of the nanodiamond can occur from the top (as in the schematic) or from the waveguide itself.}
    \label{fig:ND_IWE_scheme}
\end{figure}

The recent results developed in this article put the bricks down towards a new integrated photonics platform based on nanophotonics. Figure~\ref{fig:ND_IWE_scheme}
%TODO: put the right reference
presents a schematic of what a final structure would look like. We see a nanodiamond excited from the top and with its fluorescence coupled into the ion-exchange waveguide (side view). The photons can then be coupled to an optical fiber either using butt-coupling or by proper connectic between the fiber and the waveguide. Several issues remain to be dealt with in order to optimize this system such as an efficient interface between the nanoemitter and the waveguide. Different approaches are being explored such as the use of a plasmonic nanoantenna~\cite{apuzzo2013observation} or the deposition of a thin layer a proper-matched refractive index layer \cite{madrigal2016hybrid} or a combination of both. One this is taken care of, we still need to deal with the positioning of the nanoemitter at the right place on the waveguide which as a certain width between 1 to 2 µm. This crucial step is also investigated with different approaches such as the placing of the emitters using surface functionalisation or using an AFM tip~\cite{cuche2009near}. In the near future, it would probably be even more benefic to use a very thin diamond membrane to put directly on the waveguide for better coupling.

\section{Conclusion \& perspectives}
In this article, we presented two nanophotonics platform for efficient coupling of light with nanoemitters. We also presented recent novel results on the optical properties of perovskite nanocrystals, including quantum optical properties, as well as some results on the effect of growth and treatment on nanodiamonds monitored by the fluorescence of SiV centers. We finally explored a way towards a fully intergrated platform where light from nanoemitters is emitted directly into an optical fiber. This opens up new possibilities such as the use of the ion-exchange waveguide and nanofibers as optical buses between 2 or more nanoemitters. The potential scalability of our nanophotonic platforms is clear to us but many developements remain to be done.

\section*{Acknowledgements}
The authors would like to thank Régis Deturche, Aurélie Broussier and Xiaolun Xu from the UTT and Robert Taylor from the University of Oxford for their assitance. The authors would like to acknowledge the funding from the COST Action Nanoscale Quantum Optics, from the ITN project LIMQUET as well as the ANR projects SINPHONIE and Iper-nano2. AB et QG are members of the "Institut Universitaire de France" (IUF).

\clearpage

\bibliography{references}

\begin{thebibliography}{10}
\providecommand{\url}[1]{\texttt{#1}}
\providecommand{\urlprefix}{URL }

\bibitem{acin2018quantum}
A.~Ac{\'\i}n, I.~Bloch, H.~Buhrman, T.~Calarco, C.~Eichler, J.~Eisert,
  D.~Esteve, N.~Gisin, S.~J. Glaser, F.~Jelezko, et~al.,
\newblock \emph{New J. Phys.} \textbf{2018}, \emph{20}, 8 080201.

\bibitem{vamivakas2017special}
A.~Vamivakas, A.~Sergienko, A.~Fiore,
\newblock \emph{J. Opt.} \textbf{2017}, \emph{19}, 8 080403.

\bibitem{divincenzo2000physical}
D.~P. DiVincenzo,
\newblock \emph{Fortschritte der Physik: Progress of Physics} \textbf{2000},
  \emph{48}, 9-11 771.

\bibitem{krantz2019quantum}
P.~Krantz, M.~Kjaergaard, F.~Yan, T.~P. Orlando, S.~Gustavsson, W.~D. Oliver,
\newblock \emph{Applied Physics Reviews} \textbf{2019}, \emph{6}, 2 021318.

\bibitem{kane1998silicon}
B.~E. Kane,
\newblock \emph{Nature} \textbf{1998}, \emph{393}, 6681 133.

\bibitem{loss1998quantum}
D.~Loss, D.~P. DiVincenzo,
\newblock \emph{Phys. Rev. A} \textbf{1998}, \emph{57}, 1 120.

\bibitem{childress2013diamond}
L.~Childress, R.~Hanson,
\newblock \emph{MRS Bull.} \textbf{2013}, \emph{38}, 2 134.

\bibitem{black1988tapered}
R.~Black, E.~Gonthier, S.~Lacroix, J.~Lapierre, J.~Bures,
\newblock In \emph{Components for Fiber Optic Applications II}, volume 839.
  International Society for Optics and Photonics, \textbf{1988} 2--19.

\bibitem{tervonen_ion-exchanged_2011}
A.~Tervonen, S.~K. Honkanen, B.~R. West,
\newblock \emph{Opt. Eng.} \textbf{2011}, \emph{50}, 7 071107.

\bibitem{sipahigil2016integrated}
A.~Sipahigil, R.~E. Evans, D.~D. Sukachev, M.~J. Burek, J.~Borregaard, M.~K.
  Bhaskar, C.~T. Nguyen, J.~L. Pacheco, H.~A. Atikian, C.~Meuwly, et~al.,
\newblock \emph{Science} \textbf{2016}, \emph{354}, 6314 847.

\bibitem{stiebeiner_design_2010}
A.~Stiebeiner, R.~Garcia-Fernandez, A.~Rauschenbeutel,
\newblock \emph{Opt. Express} \textbf{2010}, \emph{18}, 22 22677.

\bibitem{nagai_ultra-low-loss_2014}
R.~Nagai, T.~Aoki,
\newblock \emph{Opt. Express} \textbf{2014}, \emph{22}, 23 28427.

\bibitem{birks_shape_1992}
T.~A. Birks, Y.~W. Li,
\newblock \emph{Journal of Lightwave Technology} \textbf{1992}, \emph{10}, 4
  432.

\bibitem{kien_field_2004}
F.~Le~Kien, J.~Liang, K.~Hakuta, V.~Balykin,
\newblock \emph{Opt. Commun.} \textbf{2004}, \emph{242}, 4-6 445.

\bibitem{nayak2008single}
K.~P. Nayak, K.~Hakuta,
\newblock \emph{New J. Phys.} \textbf{2008}, \emph{10}, 5 053003.

\bibitem{joos2018polarization}
M.~Joos, C.~Ding, V.~Loo, G.~Blanquer, E.~Giacobino, A.~Bramati,
  V.~Krachmalnicoff, Q.~Glorieux,
\newblock \emph{Phys. Rev. Appl} \textbf{2018}, \emph{9}, 6 064035.

\bibitem{hoffman2014optical}
J.~Hoffman,
\newblock Ph.D. thesis, University of Maryland, College Park, \textbf{2014}.

\bibitem{michler2000quantum}
P.~Michler, A.~Imamo{\u{g}}lu, M.~Mason, P.~Carson, G.~Strouse, S.~Buratto,
\newblock \emph{Nature} \textbf{2000}, \emph{406}, 6799 968.

\bibitem{pisanello2010room}
F.~Pisanello, L.~Martiradonna, G.~Lem{\'e}nager, P.~Spinicelli, A.~Fiore,
  L.~Manna, J.-P. Hermier, R.~Cingolani, E.~Giacobino, M.~De~Vittorio, et~al.,
\newblock \emph{Appl. Phys. Lett.} \textbf{2010}, \emph{96}, 3 033101.

\bibitem{miller1969integrated}
S.~E. Miller,
\newblock \emph{The Bell System Technical Journal} \textbf{1969}, \emph{48}, 7
  2059.

\bibitem{mazzoldi2013ion}
P.~Mazzoldi, S.~Carturan, A.~Quaranta, C.~Sada, V.~Sglavo,
\newblock \emph{Riv Nuovo Cimento} \textbf{2013}, \emph{36} 397.

\bibitem{broquin2001ion}
J.-E. Broquin,
\newblock In \emph{Integrated Optics Devices V}, volume 4277. International
  Society for Optics and Photonics, \textbf{2001} 105--117.

\bibitem{izawa1972optical}
T.~Izawa, H.~Nakagome,
\newblock \emph{Appl. Phys. Lett.} \textbf{1972}, \emph{21}, 12 584.

\bibitem{allenet2019microsensing}
T.~Allenet, F.~Geoffray, D.~Bucci, F.~Canto, P.~Moisy, J.-E. Broquin,
\newblock \emph{Opt. Eng.} \textbf{2019}, \emph{58}, 6 060502.

\bibitem{madrigal2016hybrid}
J.~B. Madrigal, R.~Tellez-Limon, F.~Gardillou, D.~Barbier, W.~Geng, C.~Couteau,
  R.~Salas-Montiel, S.~Blaize,
\newblock \emph{Appl. Opt.} \textbf{2016}, \emph{55}, 36 10263.

\bibitem{goubet2018terahertz}
N.~Goubet, A.~Jagtap, C.~Livache, B.~Martinez, H.~Portal{\`e}s, X.~Z. Xu, R.~P.
  Lobo, B.~Dubertret, E.~Lhuillier,
\newblock \emph{J. Am. Chem. Soc.} \textbf{2018}, \emph{140}, 15 5033.

\bibitem{murray1993synthesis}
C.~B. Murray, D.~J. Norris, M.~G. Bawendi,
\newblock \emph{J. Am. Chem. Soc.} \textbf{1993}, \emph{115}, 19 8706.

\bibitem{brokmann2004highly}
X.~Brokmann, E.~Giacobino, M.~Dahan, J.-P. Hermier,
\newblock \emph{Appl. Phys. Lett.} \textbf{2004}, \emph{85}, 5 712.

\bibitem{Protesescu2015}
L.~Protesescu, S.~Yakunin, M.~I. Bodnarchuk, F.~Krieg, R.~Caputo, C.~H. Hendon,
  R.~X. Yang, A.~Walsh, M.~V. Kovalenko,
\newblock \emph{Nano Lett.} \textbf{2015}, \emph{15}, 6 3692.

\bibitem{raino2019underestimated}
G.~Raino, A.~Landuyt, F.~Krieg, C.~Bernasconi, S.~T. Ochsenbein, D.~N. Dirin,
  M.~I. Bodnarchuk, M.~V. Kovalenko,
\newblock \emph{Nano Lett.} \textbf{2019}.

\bibitem{berry1977measurement}
H.~G. Berry, G.~Gabrielse, A.~Livingston,
\newblock \emph{Appl. Opt.} \textbf{1977}, \emph{16}, 12 3200.

\bibitem{manceau2018cdse}
M.~Manceau, S.~Vezzoli, Q.~Glorieux, E.~Giacobino, L.~Carbone, M.~De~Vittorio,
  J.-P. Hermier, A.~Bramati,
\newblock \emph{ChemPhysChem} \textbf{2018}, \emph{19}, 23 3288.

\bibitem{ulrich2005correlated}
S.~Ulrich, M.~Benyoucef, P.~Michler, N.~Baer, P.~Gartner, F.~Jahnke, M.~Schwab,
  H.~Kurtze, M.~Bayer, S.~Fafard, et~al.,
\newblock \emph{Phys. Rev. B} \textbf{2005}, \emph{71}, 23 235328.

\bibitem{neu2011single}
E.~Neu, D.~Steinmetz, J.~Riedrich-M{\"o}ller, S.~Gsell, M.~Fischer, M.~Schreck,
  C.~Becher,
\newblock \emph{New J. Phys.} \textbf{2011}, \emph{13}, 2 025012.

\bibitem{sternschulte19941}
H.~Sternschulte, K.~Thonke, R.~Sauer, P.~M{\"u}nzinger, P.~Michler,
\newblock \emph{Phys. Rev. B} \textbf{1994}, \emph{50}, 19 14554.

\bibitem{rogers2014multiple}
L.~J. Rogers, K.~D. Jahnke, T.~Teraji, L.~Marseglia, C.~M{\"u}ller,
  B.~Naydenov, H.~Schauffert, C.~Kranz, J.~Isoya, L.~P. McGuinness, et~al.,
\newblock \emph{Nat. Commun.} \textbf{2014}, \emph{5} 4739.

\bibitem{prawer2014quantum}
S.~Prawer, I.~Aharonovich,
\newblock \emph{Quantum information processing with diamond: Principles and
  applications},
\newblock Elsevier, \textbf{2014}.

\bibitem{su2009high}
C.-H. Su, A.~D. Greentree, L.~C. Hollenberg,
\newblock \emph{Phys. Rev. A} \textbf{2009}, \emph{80}, 5 052308.

\bibitem{davydov2014production}
V.~A. Davydov, A.~Rakhmanina, S.~Lyapin, I.~Ilichev, K.~N. Boldyrev,
  A.~Shiryaev, V.~Agafonov,
\newblock \emph{JETP Lett.} \textbf{2014}, \emph{99}, 10 585.

\bibitem{rogers2019single}
L.~J. Rogers, O.~Wang, Y.~Liu, L.~Antoniuk, C.~Osterkamp, V.~A. Davydov, V.~N.
  Agafonov, A.~B. Filipovski, F.~Jelezko, A.~Kubanek,
\newblock \emph{Phys. Rev. Appl} \textbf{2019}, \emph{11}, 2 024073.

\bibitem{neu2013low}
E.~Neu, C.~Hepp, M.~Hauschild, S.~Gsell, M.~Fischer, H.~Sternschulte,
  D.~Steinm{\"u}ller-Nethl, M.~Schreck, C.~Becher,
\newblock \emph{New J. Phys.} \textbf{2013}, \emph{15}, 4 043005.

\bibitem{apuzzo2013observation}
A.~Apuzzo, M.~F{\'e}vrier, R.~Salas-Montiel, A.~Bruyant, A.~Chelnokov,
  G.~L{\'e}rondel, B.~Dagens, S.~Blaize,
\newblock \emph{Nano Lett.} \textbf{2013}, \emph{13}, 3 1000.

\bibitem{cuche2009near}
A.~Cuche, A.~Drezet, Y.~Sonnefraud, O.~Faklaris, F.~Treussart, J.-F. Roch,
  S.~Huant,
\newblock \emph{Opt. Express} \textbf{2009}, \emph{17}, 22 19969.

\end{thebibliography}

\end{document}